\documentstyle[aps,twocolumn,psfig]{revtex}
\newcommand{\ndf}  {$\rm Nd_2 Fe_{14} B$}

\title{Ferromagnetic Resonance Force Microscopy on Microscopic Cobalt Single
Layer Films}
\author{Z. Zhang\protect\cite{pradd} and P. C. Hammel\protect\cite{corresp1}}
\address{Los Alamos National Laboratory, Los Alamos, NM 87545}
\author{M. Midzor and M. L. Roukes}
\address{California Institute of Technology, Pasadena, CA 91125}  
\author{J. R. Childress\protect\cite{pradd2}}
\address{University of Florida, Gainesville, FL 32611}
\address{
\parbox{14cm}{\bigskip\rm\small
We report mechanical detection of ferromagnetic resonance signals from 
microscopic Co single layer thin films using a magnetic resonance force microscope (MRFM).  
Variations in the magnetic anisotropy field and the inhomogeneity of 
were clearly observed in the FMR spectra of 
microscopic Co thin films 500 and 1000 \AA\ thick and  
$\sim 40 \times 200 \, \mu \mbox{m} ^2$ in lateral extent. 
This demonstrates the important potential that MRFM detection of FMR holds 
for microscopic characterization of spatial distribution of magnetic 
properties in magnetic layered materials and devices.
}}
\begin{document}
\maketitle

\thispagestyle{myheadings}
\markright{{\em LA-UR-97-5168}\hspace{30mm}{\small {\em Applied Physics Letters,} to be published October 5, 1998} \hfill }
%\pacs{PACS numbers: 76.30.-v, 76.90.+d}
\narrowtext

Ferromagnetic resonance (FMR) is an important tool for characterizing 
magnetic materials\cite{bw:fmr,heinrich:ult2}. 
It has played a particularly important role in studies of the magnetic multilayer
systems which are becoming widely used in the recording industry as 
recording read heads and/or media where    
it has been used to measure the dependence of 
the interlayer exchange coupling on the thickness of the spacer layer.  
The ability to perform microscopic FMR would be extremely valuable,
enabling characterization, on a microscopic scale, of distributions of
magnetic anisotropy and exchange energies in magnetic devices.
Microscopic FMR cannot be performed using conventional techniques 
for two reasons. 
First, the sensitivity of conventional FMR is inadequate; 
for most magnetic thin films, such as Co and Fe, sample areas on the
order of (mm)$^2$ are needed in order to obtain sufficient signal at X-band. 
Second, conventional FMR is performed in a uniform magnetic field
so there is no means to identify the spatial origin of a 
particular contribution to the FMR signal.

We have recently demonstrated the feasibility of microscopic FMR\cite{zhw} 
by means of the highly sensitive 
magnetic resonance force microscope (MRFM)\cite{r:esr,r:nmr}.
The MRFM mechanically detects the resonance signal by sensitively detecting 
the oscillatory response of a micro-mechanical resonator\cite{s:apl,s:prl}. 
Even before optimizing the instrument, 
the excellent force sensitivity of the MRFM has enabled detection of 
ferromagnetic resonance signals from magnetic thin films 
only tens of microns in lateral extent\cite{zhw}. 
As in magnetic resonance imaging, the MRFM experiment is performed 
in a strong field gradient; 
this allows the MRFM to confine the excitation of spin precession
to a well defined surface of constant magnetic field 
(or ``sensitive slice'') where the magnetic resonance condition is met. 
Spatially resolved microscopy is achieved by
scanning this sensitive slice through the sample\cite{zr:jap,hzmr}. 

In our previous demonstration of microscopic FMR using the MRFM, 
a single crystal yttrium iron garnet (YIG) thin film was used\cite{zhw}.
YIG was chosen because it has very strong FMR intensity and narrow resonance
linewidth ($<1$ Gauss). 
However, typical magnetic devices 
are composed of metallic ferromagnets such as Co and Fe.  
These have much larger FMR linewidths (of order 100 Gauss) 
and hence weaker signals, making the signals harder to detect.  
It is also essential that the resonant field be substantially larger than 
the linewidth, hence that the irradiation frequency exceed $\sim 2$--3 GHz. 
Here we report successful detection of the FMR 
signal from a microscopic, single layer Co thin film 
by means of an MRFM instrument. 
Irradiation at high frequency ($\sim 8 \,$GHz here)
and implementation of a novel MRFM geometry, the perpendicular geometry, 
as described immediately below 
were essential features of the experiment. 

The MRFM apparatus used in this work has been described earlier 
in Refs.~\onlinecite{zhw} and ~\onlinecite{hzmr}.
The polycrystalline Co films were deposited onto the 
tip of the single crystal Si cantilever over polycrystalline Ag underlayers;
the films were capped with protective Ag films. 
The sizes and locations of the films were determined by a simple mask. 
Results from three samples will be presented.
Sample 1, 1000 \AA\ thick, was sputter deposited at a substrate 
temperature $<50 ^{\circ}$C onto a 30 \AA\ Ag layer. 
Samples 2 (1000 \AA) and 3 (500 \AA) were thermally evaporated onto 
50 \AA\ Ag layers.  
A \ndf\ bar magnet ($\sim 6.4\,$mm in diameter and
$\sim 6.4 \,$mm long) was used to generate the 
field gradient $\nabla${\bf B}, and therefore a force {\bf F} on the cantilever:
\begin{equation}
F_{x} = m_{x} \frac {\partial B^{\rm bar}_{x}} {\partial x} + 
        m_{z} \frac {\partial B^{\rm bar}_{x}} {\partial z} \; ,
\label{eq:force}
\end{equation}
where the oscillatory displacement of the cantilever is parallel to the $x$-axis, 
and the axis of the bar magnet is parallel to the $z$-axis. 
The components of the magnetic moment of the Co film along these axes 
are represented by $m_{x}$ and $m_{z}$.  

\begin{figure}%[tbp]
\parbox{1mm}{\rule{0mm}{15mm}}
\parbox{3in}{\psfig{file=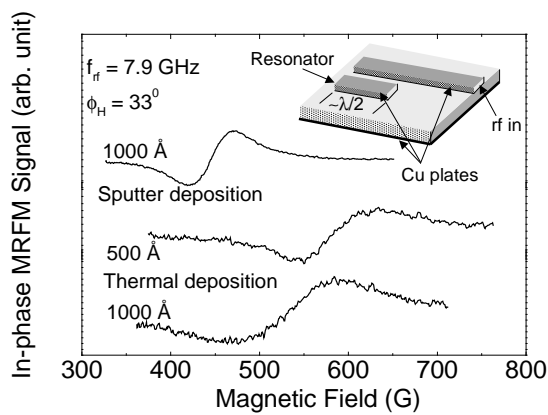,width=2.8in}}\vspace{2mm}
\caption{MRFM spectra of three single layer Co thin films using
transverse geometry MRFM.  
The experiment was performed in air and at room temperature using 
anharmonic modulation.
The angle between the external field and the film plane is 
$\phi _H \sim 33 ^{\circ}$.
The inset shows the design of the microstrip resonator.}
\label{fig:singleco}
\end{figure}

\begin{figure}
\parbox{1mm}{\rule{0mm}{15mm}}
\parbox{3in}{\psfig{file=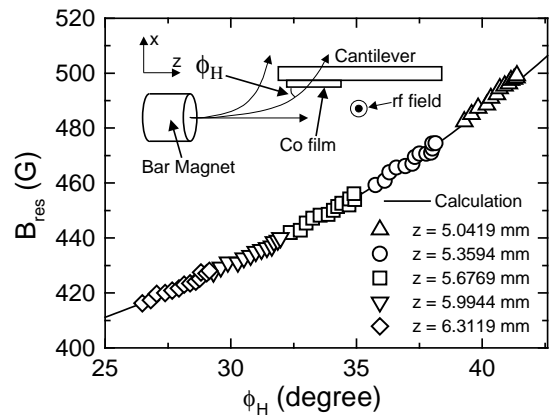,width=2.8in}}\vspace{2mm}
\caption{Dependence of the resonance field on the angle
$\phi  _H$ between the applied magnetic field and the film plane for the
sputtered sample. 
The solid line is a theoretical fit from a classical FMR
theory with $g = 2.18$ and $4\pi M_{\rm eff}= 17.6\,$kG. 
A vanishing crystalline anisotropy field was assumed in the calculation.
The inset shows a schematic diagram of the MRFM apparatus.}
\label{fig:anglefmr}
\end{figure}

For these measurements the earlier apparatus was modified 
in two respects:
we employed a geometry in which the axis 
of the bar magnet is oriented perpendicular to the motion of the cantilever, 
in contrast to the conventional MRFM geometry where these are parallel,  
and we used microstrip resonator to generate the rf field 
(see inset, Fig.\ \ref{fig:singleco}). 
The novel geometry, depicted in the inset to 
Fig.\ \ref{fig:anglefmr}, was selected because 
the applied field is in the film plane enabling us to saturate the film 
with a modest magnetic field (of order hundreds of gauss). 
In contrast, for the parallel geometry, the magnetic field would be applied
perpendicular to the film plane and $H_{\rm res} \sim 20\,$kG 
would be required in order to achieve the 
FMR condition at X-band:   
$H_{\rm res} = 4\pi M_{s} + \omega / \gamma$, 
where $\omega$ is the rf angular frequency, $\gamma$ is the gyromagnetic ratio 
and $M_{s}$ is the saturation magnetization of the film. 
The perpendicular geometry has disadvantages, however. 
First, the force in the perpendicular geometry is directly related to the field
gradient $\partial B^{\rm bar}_{x}/\partial z$ assuming that the magnetic
moment of the film stays close to the film plane. 
This reduces the sensitivity from that obtained in the parallel geometry 
where the force is proportional to a much larger gradient
$\partial B^{\rm bar}_{z}/\partial z$. 
Second, in order to obtain a finite value of 
$\partial B^{\rm bar}_{x}/\partial z$, the film must be placed off the axis
of the bar magnet. 
As a result, the external field is applied at an angle
$\phi  _H$ with respect to the film plane.
The FMR resonance field of a magnetic film varies with 
$\phi  _H$ at constant rf frequency.
Thus, a varying $\phi  _H$ complicates the interpretation of the 
experimental MRFM spectra since  
the spatial variation of both the field strength and $\phi  _H$ 
must be known in order to evaluate 
the microscopic magnetic properties of the film.

Rather than a coil (used in previous FMR experiments \cite{zhw}), 
we used a microstrip resonator with a characteristic frequency near 8 GHz
to provide the rf field. 
This increased the resonant field (applied in the film-plane), 
$H_{\rm res} = \sqrt{(2 \pi M_{s})^2 + \left( \omega / \gamma \right) ^2} 
               \; - 2 \pi M_s 
               \simeq \left(\omega / \gamma \right)^2 / 4 \pi M_s $, 
to a value sufficient to saturate the Co film ($\sim$ 100 Gauss). 
\noindent
Unlike other microstrip 
designs in which the resonator is located at the
end of the transmission line\cite{ws,r:wzwk}, the resonator 
(see inset, Fig.~\ref{fig:singleco})
is located beside the transmission line. 
This provides good coupling between the resonator and the incoming transmission 
line (-20 dB) even with a fairly large gap ($\sim 0.2 \,$mm).   

Modulation of the $z$-component of the Co spin magnetization 
was accomplished by modulating the amplitudes of both the external field $B_{0}$ 
and the rf power at two distinct frequencies whose difference 
was set equal to $f_{c}$ (i.e., anharmonic modulation)\cite{bkgs,zhw}.

Fig.~\ref{fig:singleco} shows MRFM spectra 
of the three Co single layer films.
The spectra were taken by measuring the 
in-phase oscillation amplitude of the cantilever while sweeping the
external magnetic field; 
increasing the applied field causes the position of the sensitive slice 
to move away from the source of the field gradient. 
When the sensitive slice intercepts the Co thin
film, an increase in the cantilever vibration amplitude is observed.
As in the YIG experiment\cite{zhw}, the MRFM signals
from the Co films were so large that all spectra were taken at ambient
pressure in order to reduce the $Q$-factor, and thus the response, of the cantilever
($Q \sim 15$,000 in vacuum; $\sim\,40$ in air). 
Because the signal-to-noise ratio is proportional to $\sqrt Q$, 
this result indicates that, in vacuum, the MRFM will be easily capable of 
detecting the FMR signal from Co films with similar lateral extent, 
but as thin as 20--50 \AA. 
By measuring the frequency dependence of the resonant field as a function of the
position $(x,z)$ of the sample with respect to the bar magnet, 
we distinguish the individual contributions of the electromagnet 
and the bar magnet to the total field at the film 
and determine the detailed spatial dependence of the $\hat x$ and $\hat z$ 
components of the field of the bar magnet.
From this data the angle $\phi _H$ was calculated.  
We find that the field 
\noindent
gradient $\partial B_{\rm total} / \partial z$ 
due to the bar magnet at the Co sample is $\sim 0.15 \, \mbox{G}/\mu$m. 
For our Co film, which is $\sim 200 \, \mu$m long, 
this corresponds to a 30 Gauss field difference across the film. 
Because the resonance 
linewidth of the Co
films (between 50 and 100 G) is greater than this, 
we are not able to distinguish resonance signals arising from 
different spatial locations in the microscopic Co film by means of the applied
field gradient.  
The larger field gradient produced by reducing the diameter of the magnet will be 
necessary to improve the spatial resolution of the experiment. 

Analysis of our FMR spectra enable us to determine 
magnetic properties of the microscopic ferromagnetic films;  
in particular, these spectra reveal the dependence of the magnetic anisotropy 
and the film quality on the deposition method and film thickness. 
The magnetic anisotropy of the sputtered film was determined 
by analysis of the dependence of the 
resonance field on $\phi_H$.  
This angle is varied by displacing the Co film with 
respect to the bar magnet in either the $x$ or $z$ direction, 
thus changing the $x$ and $z$ components of the magnetic field 
applied to the film. 
This result is shown in Fig.~\ref{fig:anglefmr}.
The solid curve in Fig.~\ref{fig:anglefmr} is a theoretical prediction 
for this dependence from classical FMR theory assuming only a demagnetization field. 
In particular, no crystalline anisotropy is observed; 
this absence is an expected consequence of the polycrystalline nature of the film. 
As shown, excellent agreement was obtained. 

Fig.~\ref{fig:singleco} shows the FMR spectra of the three 
films taken at constant angle $\phi _H \simeq 33 ^{\circ}$.  
The resonant field position is larger in evaporated films 
indicating an additional anisotropy in these films.  
In particular, the effective demagnetization field, 
$4 \pi M_{\rm eff}=4 \pi M_{s} - 2 K_{u} / M_{s}$ is different 
(where $K_{u}$ is the uniaxial anisotropy energy density perpendicular
to the film plane). 
This variation in $K_{u}$ could arise from different residual stresses developed 
during the film deposition.  
Assuming that the saturation magnetization $M_{s}$ for each of the Co films 
is the same as for bulk Co ($\sim$ 1400 emu/cm$^{3}$), 
we find that $2K_{u}/M_{s} \sim 0$ for the sputtered film, 
$\sim 2.8 \,$kG for the 1000 \AA\ evaporated film, 
and $\sim 4.9\,$kG for the 500 \AA\ evaporated film. 
These results indicate that the evaporated samples develop larger stress 
than the sputtered sample, 
and that the stress decreases with increasing film thickness. 

A dependence of sample homogeneity on deposition process 
is also evident in Fig.~\ref{fig:singleco} from the variation of the FMR linewidth. 
The sputtered sample has the narrowest linewidth, about 45 G.
The applied field gradient cannot explain this linewidth
variation because the sputtered sample has the largest spatial extent along
the $z$-axis (the direction along which the field gradient is largest). 
Thus, the linewidth reflects the quality or homogeneity of the Co film itself, 
and these measurement indicate that sputtering produces a more homogeneous film than 
does thermal evaporation. 
Between the two evaporated samples, 
the 500 \AA\ sample has the narrower resonance linewidth,
possibly because a two stage evaporation was required 
for the 1000 \AA\ film, 
or possibly indicating that the film quality degrades as it becomes thicker.

In conclusion, an MRFM has been successfully 
used to detect FMR signals from 
microscopic Co single layer thin films with unprecedented sensitivity.  
These signals were obtained from a perpendicular-geometry MRFM 
operating at a frequency of 8 GHz. 
These conditions ensured that, for modest resonance fields, 
the applied field component in the plane of the film was sufficient to 
saturate the film. 
The large signal intensity indicates that the sensitivity of the current
MRFM is adequate to detect FMR signals from microscopic metallic 
ferromagnets as thin as 20 \AA. 
In fact, this has been verified in our most recent experiments 
on 50 \AA\ Co films\cite{midzor}. 
Although the field gradient in the present instrument is not large enough to 
distinguish FMR signals from different spatial locations in the film, 
our results demonstrate that MRFM detection of FMR has the sensitivity to enable 
microscopic studies of systems composed of thin film metallic ferromagnets.  
We have demonstrated, in particular, the ability to observe 
variations in magnetic anisotropy energy and film quality from one
microscopic sample to another. 
In order to improve the spatial resolution, 
larger field gradients from smaller bar magnets are needed.  
Placing the magnetic probe on the cantilever is an 
essential step to enable experiments on samples 
prepared on well characterized substrates.  
These improvements are presently underway.

We gratefully acknowledge fruitful discussions with Philip Wigen at Ohio
State University and the support of the Center for Nonlinear
Studies at Los Alamos National Laboratory.  Work at Los Alamos supported by 
the United States Department of Energy, Office of Basic Energy Sciences.

%\bibliography{mrfm,MagMatls}

\begin{thebibliography}{10}

\bibitem[*]{pradd}
Present address: Seagate Technology, Mailstop 5-601, 3061 Zanker Road, San
  Jose, CA 95134.

\bibitem[\dag]{corresp1}
To whom correspondence should be addressed; e-mail: pch@lanl.gov.

\bibitem[\ddag]{pradd2}
Present address: IBM Almaden Research Center, San Jose, CA 95120.

\bibitem{bw:fmr}
P.~E. Wigen,  in {\em Magnetic Multilayers}, edited by L.~H. Bennett and R.~E.
  Watson (World Scientific Publishing Co. Pte. Ltd., Singapore, 1994), pp.\
  183--226.

\bibitem{heinrich:ult2}
B. Heinrich,  in {\em Ultrathin Magnetic Structures II}, edited by B. Heinrich
  and J.~A.~C. Bland (Springer-Verlag, Berlin Heidelberg, 1994), pp.\ 195--222.

\bibitem{zhw}
Z. Zhang, P.~C. Hammel, and P.~E. Wigen, Appl. Phys. Lett. {\bf 68},  2005
  (1996).

\bibitem{r:esr}
D. Rugar, C.~S. Yannoni, and J.~A. Sidles, Nature {\bf 360},  563  (1992).

\bibitem{r:nmr}
D. Rugar, O. Z{\"u}ger, S. T. Hoen, C. S. Yannoni, H.-M. Vieth and R. D. Kendrick, Science {\bf 264},  1560  (1994).

\bibitem{s:apl}
J.~A. Sidles, Appl. Phys. Lett. {\bf 58},  2854  (1991).

\bibitem{s:prl}
J.~A. Sidles, Phys. Rev. Lett. {\bf 68},  1124  (1992).

\bibitem{zr:jap}
O. Z{\"u}ger and D. Rugar, J. Appl. Phys. {\bf 75},  6211  (1994).

\bibitem{hzmr}
P.~C. Hammel, Z. Zhang, G.~J. Moore, and M.~L. Roukes, J. Low Temp. Phys. {\bf
  101},  59  (1995).

\bibitem{ws}
W.~J. Wallace and R.~H. Silsbee, Rev. Sci. Instr. {\bf 62},  1754  (1991).

\bibitem{r:wzwk}
K. Wago, O. Z{\"u}ger, J. Wegener, R. Kendrick, C. S. Yannoni and D. Rugar, 
Rev. Sci. Instrum. {\bf 68},  1823  (1997).

\bibitem{bkgs}
K.~J. Bruland, J. Krzystek, J.~L. Garbini, and J.~A. Sidles, Rev. Sci. Instr.
  {\bf 66},  2853  (1995).

\bibitem{midzor}
M. Midzor, M. L. Roukes, B. J. Suh, Z. Zhang, P. C. Hammel and J. R. Childress, unpublished.

\end{thebibliography}
%\bibliographystyle{prsty}

\end{document}